\documentclass[aps,prl,amsmath,amssymb,twocolumn,showpacs,superscriptaddress]{revtex4-1}
\usepackage[pdftex]{graphicx}
\usepackage[colorlinks=true,citecolor=blue]{hyperref}
\usepackage{braket}
\usepackage{graphicx,textcomp,amssymb,amsmath,dcolumn,hyperref,upgreek}

\begin{document}

\title{Optical detection of single-electron tunneling into a semiconductor quantum dot}

\author{A.~Kurzmann}
\email{annikak@phys.ethz.ch}
\affiliation{Faculty of Physics and CENIDE, University of Duisburg-Essen, Lotharstr. 1, 47057 Duisburg, Germany}
\affiliation{Solid State Physics Laboratory, ETH Zurich, 8093 Zurich, Switzerland}
\author{P.~Stegmann}
\author{J.~Kerski}
\affiliation{Faculty of Physics and CENIDE, University of Duisburg-Essen, Lotharstr. 1, 47057 Duisburg, Germany}
\author{R.~Schott}
\author{A.~Ludwig}
\author{A. D.~Wieck}
\affiliation{Chair for Applied Solid State Physics, Ruhr-Universit{\"a}t Bochum, Universit{\"a}tsstr. 150, 44780 Bochum, Germany }
\author{J.~K{\"o}nig}
\author{A.~Lorke}
\author{M.~Geller}
\affiliation{Faculty of Physics and CENIDE, University of Duisburg-Essen, Lotharstr. 1, 47057 Duisburg, Germany}
               
\date{\today}

\begin{abstract}
The maximum information of a dynamic quantum system is given by real-time detection of every quantum event, where the ultimate challenge is a stable, sensitive detector with high bandwidth. All physical information can then be drawn from a statistical analysis of the time traces. We demonstrate here an  optical detection scheme based on the time-resolved resonance fluorescence on a single quantum dot. Single-electron resolution with high signal-to-noise ratio (4$\sigma$ confidence) and high bandwidth of 10 kHz make it possible to record the individual quantum events of the transport dynamics. Full counting statistics with factorial cumulants gives access to the non-equilibrium dynamics of spin relaxation of a singly-charged dot ($\gamma_{\uparrow \downarrow} = 3$ ms$^{-1}$), even in an equilibrium transport measurement.

\end{abstract}

\maketitle


The unpredictability of a single quantum event lies at the very core of quantum mechanics. Physical information is therefore drawn from a statistical evaluation of many such processes. Nevertheless, recording each single quantum event in a time trace -- the ''random telegraph signal'' -- is of great value \cite{Blanter.2000}, as it allows insight into the underlying physical system. Here, quantum dots \cite{Petroff.2001} have proven to be well-suited systems, as they exhibit both single-photon emission and single-electron charge transport \cite{Vandersypen.2004,*Bylander.2005,*Fujisawa.2006,*Lu.2003,Marquardt.2011,*Eltrudis.2017}. While single photon emission is generally studied on self-assembled quantum dots \cite{Michler.2000,Santori.2002}, single electron transport studies are focused on gate-defined structures \cite{Gustavsson.2006,Fricke.2007,Gustavsson.2009,Flindt.2009}. We investigate, on a single self-assembled quantum dot, the single-electron transport in the optical telegraph signal with high bandwidth and observe in the full counting statistics the interplay between charge and spin dynamics in a noninvasive way. In particular, we are able to identify the {\it spin} relaxation of the Zeeman-split quantum-dot level in the {\it charge} statistics.

\begin{figure}
	\includegraphics[width=\linewidth]{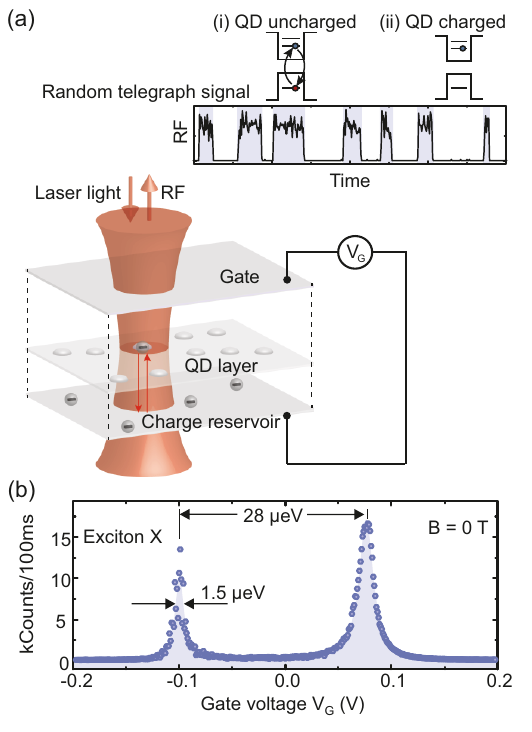}
	\caption{Optical monitoring of electron tunneling and resonance fluorescence (RF) of the QD. (a) The resonance fluorescence of the exciton transition is monitored, while an electron tunnels between the QD and a charge reservoir at a fixed gate voltage. The electron tunneling is directly observed in the time-resolved RF intensity as a random telegraph signal (depicted in the upper left inset). (b) Time-averaged resonance fluorescence of the exciton transition at zero magnetic field for a laser frequency of 325.305 THz. In a gate-voltage sweep, the fine-structure of the exciton transition is resolved with a linewidth of 1.5 $\mu$eV and a fine-structure splitting of 28 $\mu$eV.} 
	\label{fig1}
\end{figure}

The measurement of current fluctuations in conductors, semiconductors  and superconductors have provided a wealth of information about the underlying physics \cite{Blanter.2000}. Careful evaluation of shot noise, for instance, revealed the fractional charge of the quasi-particles in the fractional quantum Hall effect \cite{Kane.1994,dePicciotto.1997} and the Cooper pairing of electrons in superconductors \cite{Jehl.2000,Lefloch.2003}. Here, the ultimate challenge lies in the observation of every "quantum jump" in the so-called random telegraph signal. It represents the ultimate time resolution for a dynamic quantum system, as the complete set of time stamps, i.e. the full counting statistics \cite{Levitov.1993,*Levitov.1996}, contains the entire information about the underlying physical mechanisms. 

\begin{figure*}
	\includegraphics[width=\linewidth]{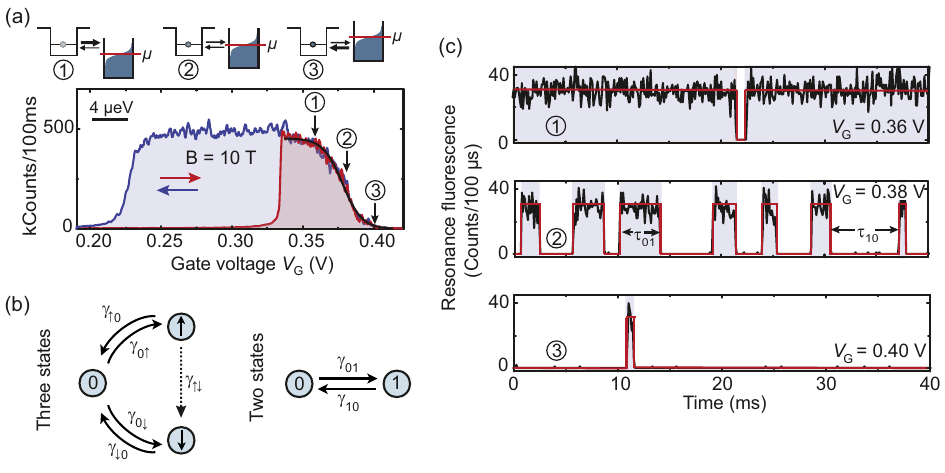}
	\caption{Time-resolved RF random telegraph signal, induced by single-electron tunneling. (a) Resonance-fluorescence intensity of the exciton transition in a magnetic field of $10\,\text{T}$ (laser frequency of 325.803 THz). Resonance is achieved over a broad range of gate voltages due to the so-called nuclear spin dragging. The Fermi distribution in the electron reservoir is observed in the RF counts between $V_\text{G}=0.33$ and $0.42\,\textrm{V}$, where the chemical potential is energetically lifted from below the QD level at point \textcircled{1} to above the QD level at point \textcircled{3}. (b) Schematic picture of the underlying model, where the quantum dot can be in three states (uncharged (0) or either charged by a spin-up ($\uparrow$) or spin-down ($\downarrow$) electron). In case of a fast spin-relaxation rate $\gamma_{\uparrow \downarrow}$, the model effectively reduces to a two-state configuration with states 0 and 1. (c) The resonance-fluorescence signal of the exciton transition measured for gate voltages \textcircled{1} -- \textcircled{3} in Fig.~\ref{fig2}(a)). The magnetic field is $B=$ 10 T, the time resolution is $100\,\upmu\textrm{s}$.}
	\label{fig2}
\end{figure*}

However, precisely recording every quantum event over long periods of time poses considerable experimental challenges, as the real-time detection needs a sensitive, low-noise and high-bandwidth detector with low back-action to a quantum system. 
For quantum dots, noninvasive voltage probes, such as a quantum point contact~\cite{Field.1993,Elzerman.2004,Vandersypen.2004}, a single-electron transistor~\cite{Lu.2003,Fujisawa.2004}, or another quantum dot \cite{Kiyama.2018} in close vicinity to the probed dot have been used as highly sensitive detectors. Counting the quantum jumps of single electrons in real-time measurements and evaluation by counting statistics yields, for instance, the degeneracy of the involved levels \cite{Hofmann.2016} or the spin-orbit interaction \cite{Maisi.2016,*Hofmann.2017}. We introduce here an optical approach, which makes it possible to record, non-invasively, the single electron dynamics of self-assembled quantum dots with single-quantum-event resolution. 

To record individual quantum events of the electron dynamics between a charge reservoir and a self-assembled dot, we use an optimized optical detection scheme based on the time-resolved resonance fluorescence on a single quantum dot \cite{Vamivakas.2010,Matthiesen.2013,Kurzmann.2016b,Kurzmann.2016c}, see Fig.~\ref{fig1}. The QD layer is embedded in a p-i-n diode structure with an highly n-doped layer as charge reservoir and a highly p-doped layer as epitaxial gate, designed for high photon out-coupling efficiency (see Fig.~\ref{fig1}(a) and Supplemental Material \cite{PRLSuppl1.2019}). The energy of the quantum-dot states with respect to the chemical potential $\mu$ of the electron reservoir is tunable by an applied gate voltage $V_\textrm{G}$, while the resonance-fluorescence signal of the exciton transition with up to 4 Mcounts/s is used as a very sensitive detector for the charging state of the quantum dot (see Supplemental Material \cite{PRLSuppl2.2019}). A typical time-averaged resonance-fluorescence signal of the exciton transition is shown in Fig.~\ref{fig1}(b). The excitation-laser frequency is 325.305 THz, and we observe the well-known doublet structure (linewidth 1.5 $\upmu$eV and fine-structure splitting of 28 $\upmu$eV) as the exciton transitions are shifted by the applied gate voltage. 

When the gate voltage further increases until eventually the 1-electron state of the quantum dot drops below the chemical potential $\mu$, a single electron can tunnel into the dot. When this happens, the transition will shift out of resonance with the laser, and the resonance-fluorescence signal will vanish. Thus, the single tunneling event can be monitored by the extinction of the optical signal (see Fig.~\ref{fig1}(a)). Setting the gate voltage so that the chemical potential in the charge reservoir is in resonance with the 1-electron state in the dot will lead to electrons randomly tunneling  between the dot and the reservoir, resulting in the random telegraph signal shown in Fig.~\ref{fig1}(a): The uncharged dot corresponds to a high RF signal (Fig.~\ref{fig1}(a)(i)), the charged dot to a low RF count (Fig.~\ref{fig1}(a)(ii)).

\begin{figure*}
	\includegraphics[width=\linewidth]{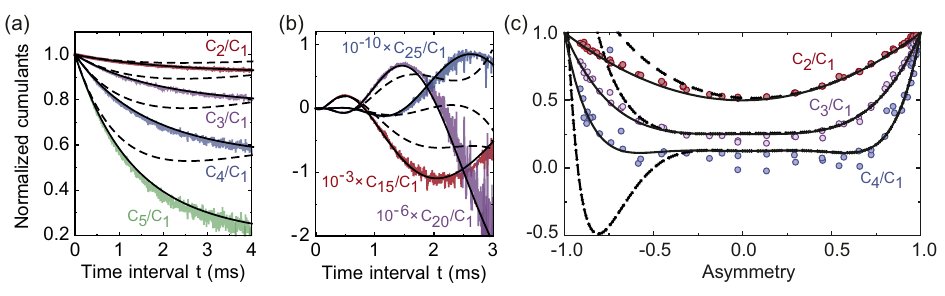}
	\caption{Analysis of the quantum-jump counting statistics. (a) -- (c) Ordinary normalized cumulants of the electron tunneling at $B=10\,\textrm{T}$ compared with simulations assuming slow (dashed) or fast (solid lines) spin relaxation. The quality of the optically obtained data allows to well resolve the cumulants as a function of time ((a),(b) with asymmetry $a=-0.9$) and asymmetry ((c) in the long-time limit for a time interval $t= 20$ ms). The dependencies, including the oscillations of the higher-order cumulants (b), are in agreement with the two-state model and clearly disagree with the three-state model, see Fig~\ref{fig2}(b).}
	\label{fig3}
\end{figure*}

Small fluctuations ($<1\,\upmu\textrm{eV}$) in the exciton transition energy, induced by spurious charges in the dot's vicinity lead to strong RF intensity noise \cite{Kuhlmann.2013}. To improve the stability of the measurement, we apply a magnetic field and make use of nuclear spin dragging, i.e. a buildup of nuclear spin polarization under resonant excitation \cite{Latta.2009,*Hogele.2012}. This will constitute an internal feedback mechanism, which pins the resonance to the fixed laser frequency and, thus, reduces the spurious charge-induced noise. Moreover, nuclear dragging also allows us to map out the Fermi distribution of the electron reservoir, by simply shifting the gate bias, without the need to readjust the laser energy to account for the bias-induced Stark shift, see Fig.~\ref{fig2}(a).   

Figure~\ref{fig2}(c) displays the optically-detected random telegraph signal for an external magnetic field of $B=$ 10 T.  At gate voltages around and below $0.36\,\textrm{V}$, the chemical potential $\mu$ of the reservoir is more than one unit of thermal energy $kT$ below the 1-electron state, so that the dot will be mostly uncharged and the time-integrated resonance-fluorescence signal will be at maximum (see vertical arrow  \textcircled{1} in Fig.~\ref{fig2}(a) and the top time-trace in Fig.~\ref{fig2}(c)). For gate voltages around $0.40\,\textrm{V}$, the chemical potential lies well above the single-electron state, the dot is therefore mostly charged and the optical signal vanishes, see vertical arrow \textcircled{3} in Fig.~\ref{fig2}(a) and the bottom time trace in Fig.~\ref{fig2}(c). In the bias range  $0.35\,\textrm{V}< V_\textrm{G}<0.4\,\textrm{V}$, the time-averaged intensity directly reflects the Fermi distribution, as seen from the fit to the data (black line in Fig.~\ref{fig2}(a)).

As seen in Fig.~\ref{fig2}(c), the on-signal (uncharged dot) and the off-signal (dot charged with a single electron) have clearly distinguishable intensities. Probability distributions of both signals are given in the Supplemental Material \cite{PRLSuppl2.2019} and show that for a threshold between 4 and 5 counts per $100\,\mu \textrm{s}$, we can distinguish between the on and off state with high confidence (4$\sigma$). The time resolution of our experiment is given by the photon-count binning of 100 $\mu$s. Recording the random telegraph signal uninterruptedly for up to 30 minutes, i.e. $2\times 10^7$ binning times, gives us a large set of data, which allows it to evaluate the tunneling dynamics in great detail.

An external magnetic field lifts the spin degeneracy of the lowest energy level in the quantum dot. 
The quantum jumps are, then, described within a three-state model, as schematically depicted in Fig.~\ref{fig2}(b), left.
Charging the quantum dot with an electron of spin $\sigma=\uparrow,\downarrow$ changes its state from 0 to $\sigma$ with rate $\gamma_{0\sigma}$.
The rate for discharging of the quantum dot with spin $\sigma$ is $\gamma_{\sigma 0}$.
Finally, spin relaxation from state $\uparrow$ to $\downarrow$ occurs with rate $\gamma_{\uparrow \downarrow}$.
Spin relaxation does not change the quantum-dot charge.
Nevertheless, its presence influences the charge-transfer statistics, which allows to infer the spin-relaxation rate by analyzing full counting statistics, as we show in the following.

Using the common procedures for analyzing full counting data sets \cite{Gustavsson.2009}, the entire time trace is divided into slices of length $t$. 
In these slices, the number $N$ of tunneling events (either in or out) is determined to obtain the probability distribution $P_N(t)$.
We then derive the cumulants $C_{m}(t)= \partial_z^m \ln {\cal M}(z,t)|_{z=0}$ from the generating function ${\cal M}(z,t)= \sum_{N=0}^\infty e^{Nz} P_N(t)$.
The first cumulant $C_1(t)$ is the mean value, $C_2(t)$ the variance, and so forth.
A selection of cumulants, ranging from $C_2(t)$ to $C_{25}(t)$ is shown in Fig.~\ref{fig3}(a) and (b) as a function of $t$ for $B=10$ T. We resolve several oscillations in the higher-order cumulants. While their rich structure pretends to convey increasing specific information about the underlying stochastic system, their appearance has been shown to be a universal feature instead. Nevertheless, we can use the higher-order cumulants to demonstrate that the quality of our data matches the most extensive data taken by transport spectroscopy \cite{Flindt.2009}. We simulate the time dependence for the extreme cases of negligible spin relaxation, $\gamma_{\uparrow \downarrow}=0$ (dashed lines) and very fast spin relaxation, $\gamma_{\uparrow \downarrow}=\infty$ (solid lines), see also Supplemental Material \cite{PRLSuppl3.2019}.
Good agreement is achieved for the latter case, for which the three-state model effectively reduces to a two-state configuration, with states 0 and 1 (see Fig.~\ref{fig2}(b))
as well as rates $\gamma_{01}=\gamma_{0 \uparrow}+\gamma_{0 \downarrow}$ for charging and $\gamma_\textrm{10}=\gamma_{\downarrow0}$ for discharging.
The good agreement holds for all asymmetries $a=(\gamma_{10}-\gamma_{01})/(\gamma_{10}+\gamma_{01})$ of the charging and discharging rates, as depicted in Fig.~\ref{fig3}(c) (solid lines) in the limit of large $t$, for which the cumulants are given by \cite{Gustavsson.2006} $C_2/C_1 = (1+a^2)/2, \, C_3/C_1 = (1+3a^4)/4$ and $C_4/C_1 = (1+a^2 -9a^4 +15a^6)/8 \, ,$
as typically found in the context of gate-defined quantum dots with fast spin-relaxation rates~\cite{Hanson.2005}.

\begin{figure*}[t]
{\includegraphics[width=\linewidth]{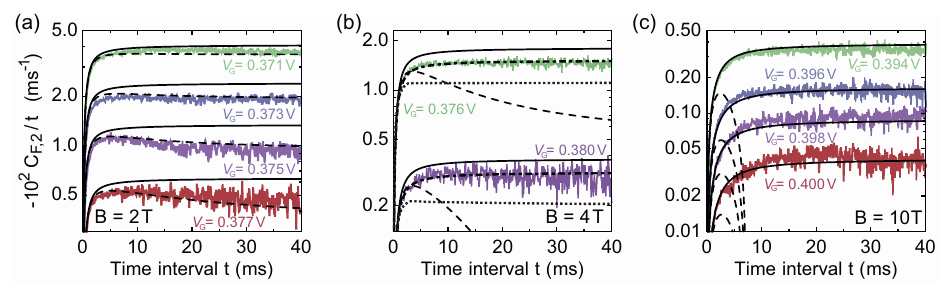}}
\caption{Interplay between charge and spin dynamics. (a) -- (c) Factorial cumulants for an applied magnetic field of $2\, \text{T}$ (a), $4\, \text{T}$ (b), and $10\, \text{T}$ (c) and excitation energies $\varepsilon/k_\text{B}T$ of $-1.7$ (green), $-2.0$ (blue), $-2.3$ (purple), and $-2.6$ (red). The black lines are obtained from simulation with spin-relaxation rate $\gamma_{\uparrow\downarrow}= \infty$ (solid), $\gamma_{\uparrow\downarrow}=3 \,\text{ms}^{-1}$ (dot-dashed), $\gamma_{\uparrow\downarrow}=1\, \text{ms}^{-1}$ (dotted), and $\gamma_{\uparrow\downarrow}=0$ (dashed). Decreasing the magnetic field yields a transition from fast to slow spin relaxation. For  $10\, \text{T}$, the data are reproduced by assuming fast spin relaxation $\gamma_{\uparrow\downarrow}= \infty$, at $4\, \text{T}$ best agreement is obtained for $\gamma_{\uparrow\downarrow}= 3\,\text{ms}^{-1}$, and at $2\, \text{T}$ we have to assume slow spin relaxation $\gamma_{\uparrow\downarrow}= 0$.}
\label{fig:figcum}
\end{figure*}

With decreasing magnetic field, the relaxation rate decreases~\cite{Kroutvar.2004} until it becomes comparable to and then smaller than the tunneling rates such that the three-state configuration can no longer be simplified to a two-state model.
We continue the analysis of our data by making use of the so-called \emph{factorial} cumulants $C_{\text{F},m}(t)= \partial_z^m \ln {\cal M}_\text{F}(z,t)|_{z=0}$ derived from ${\cal M}_\text{F}(z,t)= \sum_N (z+1)^N P_N(t)$~\cite{Kambly.2011,Stegmann.2015,*Stegmann.2016}.
While \emph{ordinary} cumulants generically change sign as a function of time and asymmetry \cite{Flindt.2009,Fricke.2010}
the \emph{factorial} cumulants of the very same probability distribution are much more well-behaved.
In fact, for a two-state system, their sign is fixed, $C_{\text{F},m}(t) \propto (-1)^{m-1}$. 
Introducing less structure into the evaluated data makes the factorial cumulants better suited to identify deviations from the two-state system.
As a second, practical advantage, higher-order factorial cumulants are more robust against imperfections of the detector than ordinary cumulants are.
Rarely occurring tunneling events yield approximately Poissonian counting statistics, for which the ordinary cumulants of all orders are equal, $C_m=C_1$, i.e., any statistical error in $C_1$ propagates to $C_m$. 
Higher-order factorial cumulants, on the other hand, measure the deviations from a Poissonian distribution (since for the latter $C_{\text{F},m}=0$ for all $m > 1$), and imperfections of the detector mainly affect $C_{\text{F},1}$ only.

In Fig.~\ref{fig:figcum}, we show the second factorial cumulant as a function of the time-interval length $t$ for different magnetic fields and different excitation energies (tunable by the gate voltage). 
The excitation energy $\varepsilon$ is defined as the difference between the mean of the Zeeman-split quantum-dot levels ($E_\uparrow$, $E_\downarrow$)  and the Fermi energy $\mu$ of the charge reservoir: $\varepsilon=(E_\uparrow+E_\downarrow)/2-\mu$.
For convenience, we divide $C_{\text{F},2}(t)$ by $t$, so that the curves approach a constant limit and do not increase linearly for long times.
The results for $10\,\text{T}$ are depicted in Fig.~\ref{fig:figcum}(c).
Again, we compare the experimental results with two simulations, one in which we assume very slow, $\gamma_{\uparrow\downarrow}=0$ (dashed) and one in which we assume fast, $\gamma_{\uparrow\downarrow}=\infty$ (solid lines) spin relaxation, see the Supplemental Material \cite{PRLSuppl3.2019}. 
We find very good agreement only for $\gamma_{\uparrow\downarrow}=\infty$. 
The simulation assuming for $\gamma_{\uparrow\downarrow}=0$ (dashed lines) is completely off.
It would even predict a sign change of the second factorial cumulant at $t < 10\, \text{ms}$, in clear contrast to the measured data.
At a magnetic field of $2 \, \text{T}$, see Fig.~\ref{fig:figcum}(a), we find the opposite situation.
Here, the experimental data are very well reproduced by assuming that the spin relaxation is so slow that it can be neglected.
For an intermediate magnetic field of $4\,\text{T}$, neither a very fast nor a very slow spin relaxation can fit the data, as can be seen in Fig.~\ref{fig:figcum}(b).
Instead, we find a spin-relaxation rate of around $\gamma_{\uparrow\downarrow}=3\, \text{ms}^{-1}$ (dashed-dot curves), which is of the same order of magnitude as the time scale for the charge dynamics.
We remark that trying to deduce $\gamma_{\uparrow \downarrow}$ from analyzing ordinary cumulants fails unless the finite time resolution of the detector is explicitly taken into account in the simulations (see Supplemental Material \cite{PRLSuppl3.2019}). Factorial cumulants do not suffer from this problem, and already the second factorial cumulant is sufficient to determine the spin-relaxation rate.

In summary, we have introduced an optical technique to record the full counting statistics of single-electron transport between a reservoir and a single self-assembled quantum dot. A high signal-to-noise ratio has been achieved with a bandwidth determined by the photon-emission rate. A bandwidth of 10 kHz is obtained for a rate of 200 kcounts/s. Photon rates above 2 Mcounts/s (see Supplemental Material \cite{PRLSuppl2.2019}) can increase the bandwidth further by one order of magnitude, outperforming electrical readout techniques for single quantum events. By analyzing the obtained random telegraph signal with recently-introduced factorial cumulants, we were able to discriminate between the statistics of a three-state system, with two distinguishable spin configurations, and the two-state analog, resulting from fast spin relaxation. By adjusting an external magnetic field, the three fundamental cases of the spin relaxation rate being much smaller, of the same order, and much larger than the tunneling rate were realized and identified by their respective statistical fingerprint. At 4 T, a spin-relaxation rate of $3\, \text{ms}^{-1}$ was found. We would like to point out that, even though the \emph{uncharged} dot was optically probed and  the sample was in transport \emph{equilibrium}, our experiments give access to a non-equilibrium property (namely the spin-relaxation rate) of the singly-charged dot.

We conclude by emphasizing that our proposed optical scheme using resonance fluorescence to probe the dynamics of a quantum system is not limited to self-assembled semiconductor quantum dots. The scheme is rather generally applicable to other types of quantum dots, active impurity states (such as vacancy or color centers) or molecular systems. Furthermore, the stochastic dynamics probed by the technique is not limited to single-electron tunneling but may include other stochastic processes such as conformational changes or magnetic excitations of molecules.

\begin{acknowledgements}
This work was supported by the German Research Foundation (DFG) within the Collaborative Research Centre (SFB) 1242 "Non-Equilibrium Dynamics of Condensed Matter in the Time Domain" (Project number 278162697) and the individual research grant No.~GE2141/5-1 and LU2051/1-1. Ar.~L.,R.~S., and A.~D.~W. acknowledge gratefully support of DFG-TRR-160, BMBF-Q.Link.X 16KIS0867, and the DFH/UFA CDFA-05-06.
\end{acknowledgements}


\begin{thebibliography}{48}%
	\makeatletter
	\providecommand \@ifxundefined [1]{%
		\@ifx{#1\undefined}
	}%
	\providecommand \@ifnum [1]{%
		\ifnum #1\expandafter \@firstoftwo
		\else \expandafter \@secondoftwo
		\fi
	}%
	\providecommand \@ifx [1]{%
		\ifx #1\expandafter \@firstoftwo
		\else \expandafter \@secondoftwo
		\fi
	}%
	\providecommand \natexlab [1]{#1}%
	\providecommand \enquote  [1]{``#1''}%
	\providecommand \bibnamefont  [1]{#1}%
	\providecommand \bibfnamefont [1]{#1}%
	\providecommand \citenamefont [1]{#1}%
	\providecommand \href@noop [0]{\@secondoftwo}%
	\providecommand \href [0]{\begingroup \@sanitize@url \@href}%
	\providecommand \@href[1]{\@@startlink{#1}\@@href}%
	\providecommand \@@href[1]{\endgroup#1\@@endlink}%
	\providecommand \@sanitize@url [0]{\catcode `\\12\catcode `\$12\catcode
		`\&12\catcode `\#12\catcode `\^12\catcode `\_12\catcode `\%12\relax}%
	\providecommand \@@startlink[1]{}%
	\providecommand \@@endlink[0]{}%
	\providecommand \url  [0]{\begingroup\@sanitize@url \@url }%
	\providecommand \@url [1]{\endgroup\@href {#1}{\urlprefix }}%
	\providecommand \urlprefix  [0]{URL }%
	\providecommand \Eprint [0]{\href }%
	\providecommand \doibase [0]{http://dx.doi.org/}%
	\providecommand \selectlanguage [0]{\@gobble}%
	\providecommand \bibinfo  [0]{\@secondoftwo}%
	\providecommand \bibfield  [0]{\@secondoftwo}%
	\providecommand \translation [1]{[#1]}%
	\providecommand \BibitemOpen [0]{}%
	\providecommand \bibitemStop [0]{}%
	\providecommand \bibitemNoStop [0]{.\EOS\space}%
	\providecommand \EOS [0]{\spacefactor3000\relax}%
	\providecommand \BibitemShut  [1]{\csname bibitem#1\endcsname}%
	\let\auto@bib@innerbib\@empty
	\bibitem [{\citenamefont {Blanter}\ and\ \citenamefont
		{B{\"u}ttiker}(2000)}]{Blanter.2000}%
	\BibitemOpen
	\bibfield  {author} {\bibinfo {author} {\bibfnamefont {Y.~M.}\ \bibnamefont
			{Blanter}}\ and\ \bibinfo {author} {\bibfnamefont {M.}~\bibnamefont
			{B{\"u}ttiker}},\ }\href {\doibase 10.1016/S0370-1573(99)00123-4} {\bibfield
		{journal} {\bibinfo  {journal} {Physics Reports}\ }\textbf {\bibinfo {volume}
			{336}},\ \bibinfo {pages} {1} (\bibinfo {year} {2000})}\BibitemShut {NoStop}%
	\bibitem [{\citenamefont {Petroff}\ \emph {et~al.}(2001)\citenamefont
		{Petroff}, \citenamefont {Lorke},\ and\ \citenamefont
		{Imamoglu}}]{Petroff.2001}%
	\BibitemOpen
	\bibfield  {author} {\bibinfo {author} {\bibfnamefont {P.~M.}\ \bibnamefont
			{Petroff}}, \bibinfo {author} {\bibfnamefont {A.}~\bibnamefont {Lorke}}, \
		and\ \bibinfo {author} {\bibfnamefont {A.}~\bibnamefont {Imamoglu}},\
	}\href@noop {} {\bibfield  {journal} {\bibinfo  {journal} {Physics Today}\
		}\textbf {\bibinfo {volume} {54}},\ \bibinfo {pages} {46} (\bibinfo {year}
		{2001})}\BibitemShut {NoStop}%
	\bibitem [{\citenamefont {Vandersypen}\ \emph {et~al.}(2004)\citenamefont
		{Vandersypen}, \citenamefont {Elzerman}, \citenamefont {Schouten},
		\citenamefont {{Beveren, L. H. Willems van Beveren}}, \citenamefont
		{Hanson},\ and\ \citenamefont {Kouwenhoven}}]{Vandersypen.2004}%
	\BibitemOpen
	\bibfield  {author} {\bibinfo {author} {\bibfnamefont {L.~M.~K.}\
			\bibnamefont {Vandersypen}}, \bibinfo {author} {\bibfnamefont {J.~M.}\
			\bibnamefont {Elzerman}}, \bibinfo {author} {\bibfnamefont {R.~N.}\
			\bibnamefont {Schouten}}, \bibinfo {author} {\bibnamefont {{Beveren, L. H.
					Willems van Beveren}}}, \bibinfo {author} {\bibfnamefont {R.}~\bibnamefont
			{Hanson}}, \ and\ \bibinfo {author} {\bibfnamefont {L.~P.}\ \bibnamefont
			{Kouwenhoven}},\ }\href {\doibase 10.1063/1.1815041} {\bibfield  {journal}
		{\bibinfo  {journal} {Appl. Phys. Lett.}\ }\textbf {\bibinfo {volume} {85}},\
		\bibinfo {pages} {4394} (\bibinfo {year} {2004})}\BibitemShut {NoStop}%
	\bibitem [{\citenamefont {Bylander}\ \emph {et~al.}(2005)\citenamefont
		{Bylander}, \citenamefont {Duty},\ and\ \citenamefont
		{Delsing}}]{Bylander.2005}%
	\BibitemOpen
	\bibfield  {author} {\bibinfo {author} {\bibfnamefont {J.}~\bibnamefont
			{Bylander}}, \bibinfo {author} {\bibfnamefont {T.}~\bibnamefont {Duty}}, \
		and\ \bibinfo {author} {\bibfnamefont {P.}~\bibnamefont {Delsing}},\ }\href
	{\doibase 10.1038/nature03375} {\bibfield  {journal} {\bibinfo  {journal}
			{Nature}\ }\textbf {\bibinfo {volume} {434}},\ \bibinfo {pages} {361}
		(\bibinfo {year} {2005})}\BibitemShut {NoStop}%
	\bibitem [{\citenamefont {Fujisawa}\ \emph {et~al.}(2006)\citenamefont
		{Fujisawa}, \citenamefont {Hayashi}, \citenamefont {Tomita},\ and\
		\citenamefont {Hirayama}}]{Fujisawa.2006}%
	\BibitemOpen
	\bibfield  {author} {\bibinfo {author} {\bibfnamefont {T.}~\bibnamefont
			{Fujisawa}}, \bibinfo {author} {\bibfnamefont {T.}~\bibnamefont {Hayashi}},
		\bibinfo {author} {\bibfnamefont {R.}~\bibnamefont {Tomita}}, \ and\ \bibinfo
		{author} {\bibfnamefont {Y.}~\bibnamefont {Hirayama}},\ }\href {\doibase
		10.1126/science.1126788} {\bibfield  {journal} {\bibinfo  {journal}
			{Science}\ }\textbf {\bibinfo {volume} {312}},\ \bibinfo {pages} {1634}
		(\bibinfo {year} {2006})}\BibitemShut {NoStop}%
	\bibitem [{\citenamefont {Lu}\ \emph {et~al.}(2003)\citenamefont {Lu},
		\citenamefont {Ji}, \citenamefont {Pfeiffer}, \citenamefont {West},\ and\
		\citenamefont {Rimberg}}]{Lu.2003}%
	\BibitemOpen
	\bibfield  {author} {\bibinfo {author} {\bibfnamefont {W.}~\bibnamefont
			{Lu}}, \bibinfo {author} {\bibfnamefont {Z.}~\bibnamefont {Ji}}, \bibinfo
		{author} {\bibfnamefont {L.}~\bibnamefont {Pfeiffer}}, \bibinfo {author}
		{\bibfnamefont {K.~W.}\ \bibnamefont {West}}, \ and\ \bibinfo {author}
		{\bibfnamefont {A.~J.}\ \bibnamefont {Rimberg}},\ }\href {\doibase
		10.1038/nature01642} {\bibfield  {journal} {\bibinfo  {journal} {Nature}\
		}\textbf {\bibinfo {volume} {423}},\ \bibinfo {pages} {422} (\bibinfo {year}
		{2003})}\BibitemShut {NoStop}%
	\bibitem [{\citenamefont {Marquardt}\ \emph {et~al.}(2011)\citenamefont
		{Marquardt}, \citenamefont {Geller}, \citenamefont {Baxevanis}, \citenamefont
		{Pfannkuche}, \citenamefont {Wieck}, \citenamefont {Reuter},\ and\
		\citenamefont {Lorke}}]{Marquardt.2011}%
	\BibitemOpen
	\bibfield  {author} {\bibinfo {author} {\bibfnamefont {B.}~\bibnamefont
			{Marquardt}}, \bibinfo {author} {\bibfnamefont {M.}~\bibnamefont {Geller}},
		\bibinfo {author} {\bibfnamefont {B.}~\bibnamefont {Baxevanis}}, \bibinfo
		{author} {\bibfnamefont {D.}~\bibnamefont {Pfannkuche}}, \bibinfo {author}
		{\bibfnamefont {A.~D.}\ \bibnamefont {Wieck}}, \bibinfo {author}
		{\bibfnamefont {D.}~\bibnamefont {Reuter}}, \ and\ \bibinfo {author}
		{\bibfnamefont {A.}~\bibnamefont {Lorke}},\ }\href {\doibase
		10.1038/ncomms1205} {\bibfield  {journal} {\bibinfo  {journal} {Nature
				Commun.}\ }\textbf {\bibinfo {volume} {2}},\ \bibinfo {pages} {209} (\bibinfo
		{year} {2011})}\BibitemShut {NoStop}%
	\bibitem [{\citenamefont {Eltrudis}\ \emph {et~al.}(2017)\citenamefont
		{Eltrudis}, \citenamefont {Al-Ashouri}, \citenamefont {Beckel}, \citenamefont
		{Ludwig}, \citenamefont {Wieck}, \citenamefont {Geller},\ and\ \citenamefont
		{Lorke}}]{Eltrudis.2017}%
	\BibitemOpen
	\bibfield  {author} {\bibinfo {author} {\bibfnamefont {K.}~\bibnamefont
			{Eltrudis}}, \bibinfo {author} {\bibfnamefont {A.}~\bibnamefont
			{Al-Ashouri}}, \bibinfo {author} {\bibfnamefont {A.}~\bibnamefont {Beckel}},
		\bibinfo {author} {\bibfnamefont {A.}~\bibnamefont {Ludwig}}, \bibinfo
		{author} {\bibfnamefont {A.~D.}\ \bibnamefont {Wieck}}, \bibinfo {author}
		{\bibfnamefont {M.}~\bibnamefont {Geller}}, \ and\ \bibinfo {author}
		{\bibfnamefont {A.}~\bibnamefont {Lorke}},\ }\href {\doibase
		10.1063/1.4985572} {\bibfield  {journal} {\bibinfo  {journal} {Appl. Phys.
				Lett.}\ }\textbf {\bibinfo {volume} {111}},\ \bibinfo {pages} {092103}
		(\bibinfo {year} {2017})}\BibitemShut {NoStop}%
	\bibitem [{\citenamefont {Michler}\ \emph {et~al.}(2000)\citenamefont
		{Michler}, \citenamefont {Kiraz}, \citenamefont {Zhang}, \citenamefont
		{Becher}, \citenamefont {Hu},\ and\ \citenamefont {Imamoglu}}]{Michler.2000}%
	\BibitemOpen
	\bibfield  {author} {\bibinfo {author} {\bibfnamefont {P.}~\bibnamefont
			{Michler}}, \bibinfo {author} {\bibfnamefont {A.}~\bibnamefont {Kiraz}},
		\bibinfo {author} {\bibfnamefont {L.}~\bibnamefont {Zhang}}, \bibinfo
		{author} {\bibfnamefont {C.}~\bibnamefont {Becher}}, \bibinfo {author}
		{\bibfnamefont {E.}~\bibnamefont {Hu}}, \ and\ \bibinfo {author}
		{\bibfnamefont {A.}~\bibnamefont {Imamoglu}},\ }\href {\doibase
		10.1063/1.126918} {\bibfield  {journal} {\bibinfo  {journal} {Appl. Phys.
				Lett.}\ }\textbf {\bibinfo {volume} {77}},\ \bibinfo {pages} {184} (\bibinfo
		{year} {2000})}\BibitemShut {NoStop}%
	\bibitem [{\citenamefont {Santori}\ \emph {et~al.}(2002)\citenamefont
		{Santori}, \citenamefont {Fattal}, \citenamefont {Vuckovi{\'c}},
		\citenamefont {Solomon},\ and\ \citenamefont {Yamamoto}}]{Santori.2002}%
	\BibitemOpen
	\bibfield  {author} {\bibinfo {author} {\bibfnamefont {C.}~\bibnamefont
			{Santori}}, \bibinfo {author} {\bibfnamefont {D.}~\bibnamefont {Fattal}},
		\bibinfo {author} {\bibfnamefont {J.}~\bibnamefont {Vuckovi{\'c}}}, \bibinfo
		{author} {\bibfnamefont {G.~S.}\ \bibnamefont {Solomon}}, \ and\ \bibinfo
		{author} {\bibfnamefont {Y.}~\bibnamefont {Yamamoto}},\ }\href {\doibase
		10.1109/QELS.2003.1276041} {\bibfield  {journal} {\bibinfo  {journal}
			{Nature}\ }\textbf {\bibinfo {volume} {419}},\ \bibinfo {pages} {594}
		(\bibinfo {year} {2002})}\BibitemShut {NoStop}%
	\bibitem [{\citenamefont {Gustavsson}\ \emph {et~al.}(2006)\citenamefont
		{Gustavsson}, \citenamefont {Leturcq}, \citenamefont {Simovi{\v{c}}},
		\citenamefont {Schleser}, \citenamefont {Ihn}, \citenamefont {Studerus},
		\citenamefont {Ensslin}, \citenamefont {Driscoll},\ and\ \citenamefont
		{Gossard}}]{Gustavsson.2006}%
	\BibitemOpen
	\bibfield  {author} {\bibinfo {author} {\bibfnamefont {S.}~\bibnamefont
			{Gustavsson}}, \bibinfo {author} {\bibfnamefont {R.}~\bibnamefont {Leturcq}},
		\bibinfo {author} {\bibfnamefont {B.}~\bibnamefont {Simovi{\v{c}}}}, \bibinfo
		{author} {\bibfnamefont {R.}~\bibnamefont {Schleser}}, \bibinfo {author}
		{\bibfnamefont {T.}~\bibnamefont {Ihn}}, \bibinfo {author} {\bibfnamefont
			{P.}~\bibnamefont {Studerus}}, \bibinfo {author} {\bibfnamefont
			{K.}~\bibnamefont {Ensslin}}, \bibinfo {author} {\bibfnamefont {D.~C.}\
			\bibnamefont {Driscoll}}, \ and\ \bibinfo {author} {\bibfnamefont {A.~C.}\
			\bibnamefont {Gossard}},\ }\href {\doibase 10.1103/PhysRevLett.96.076605}
	{\bibfield  {journal} {\bibinfo  {journal} {Phys. Rev. Lett.}\ }\textbf
		{\bibinfo {volume} {96}},\ \bibinfo {pages} {076605} (\bibinfo {year}
		{2006})}\BibitemShut {NoStop}%
	\bibitem [{\citenamefont {Fricke}\ \emph {et~al.}(2007)\citenamefont {Fricke},
		\citenamefont {Hohls}, \citenamefont {Wegscheider},\ and\ \citenamefont
		{Haug}}]{Fricke.2007}%
	\BibitemOpen
	\bibfield  {author} {\bibinfo {author} {\bibfnamefont {C.}~\bibnamefont
			{Fricke}}, \bibinfo {author} {\bibfnamefont {F.}~\bibnamefont {Hohls}},
		\bibinfo {author} {\bibfnamefont {W.}~\bibnamefont {Wegscheider}}, \ and\
		\bibinfo {author} {\bibfnamefont {R.~J.}\ \bibnamefont {Haug}},\ }\href
	{\doibase 10.1103/PhysRevB.76.155307} {\bibfield  {journal} {\bibinfo
			{journal} {Phys. Rev. B}\ }\textbf {\bibinfo {volume} {76}},\ \bibinfo
		{pages} {155307} (\bibinfo {year} {2007})}\BibitemShut {NoStop}%
	\bibitem [{\citenamefont {Gustavsson}\ \emph {et~al.}(2009)\citenamefont
		{Gustavsson}, \citenamefont {Leturcq}, \citenamefont {Studer}, \citenamefont
		{Shorubalko}, \citenamefont {Ihn}, \citenamefont {Ensslin}, \citenamefont
		{Driscoll},\ and\ \citenamefont {Gossard}}]{Gustavsson.2009}%
	\BibitemOpen
	\bibfield  {author} {\bibinfo {author} {\bibfnamefont {S.}~\bibnamefont
			{Gustavsson}}, \bibinfo {author} {\bibfnamefont {R.}~\bibnamefont {Leturcq}},
		\bibinfo {author} {\bibfnamefont {M.}~\bibnamefont {Studer}}, \bibinfo
		{author} {\bibfnamefont {I.}~\bibnamefont {Shorubalko}}, \bibinfo {author}
		{\bibfnamefont {T.}~\bibnamefont {Ihn}}, \bibinfo {author} {\bibfnamefont
			{K.}~\bibnamefont {Ensslin}}, \bibinfo {author} {\bibfnamefont {D.~C.}\
			\bibnamefont {Driscoll}}, \ and\ \bibinfo {author} {\bibfnamefont {A.~C.}\
			\bibnamefont {Gossard}},\ }\href {\doibase 10.1016/j.surfrep.2009.02.001}
	{\bibfield  {journal} {\bibinfo  {journal} {Surf. Sci. Rep.}\ }\textbf
		{\bibinfo {volume} {64}},\ \bibinfo {pages} {191} (\bibinfo {year}
		{2009})}\BibitemShut {NoStop}%
	\bibitem [{\citenamefont {Flindt}\ \emph {et~al.}(2009)\citenamefont {Flindt},
		\citenamefont {Fricke}, \citenamefont {Hohls}, \citenamefont {Novotn{\'y}},
		\citenamefont {Neto{\v{c}}n{\'y}}, \citenamefont {Brandes},\ and\
		\citenamefont {Haug}}]{Flindt.2009}%
	\BibitemOpen
	\bibfield  {author} {\bibinfo {author} {\bibfnamefont {C.}~\bibnamefont
			{Flindt}}, \bibinfo {author} {\bibfnamefont {C.}~\bibnamefont {Fricke}},
		\bibinfo {author} {\bibfnamefont {F.}~\bibnamefont {Hohls}}, \bibinfo
		{author} {\bibfnamefont {T.}~\bibnamefont {Novotn{\'y}}}, \bibinfo {author}
		{\bibfnamefont {K.}~\bibnamefont {Neto{\v{c}}n{\'y}}}, \bibinfo {author}
		{\bibfnamefont {T.}~\bibnamefont {Brandes}}, \ and\ \bibinfo {author}
		{\bibfnamefont {R.~J.}\ \bibnamefont {Haug}},\ }\href {\doibase
		10.1073/pnas.0901002106} {\bibfield  {journal} {\bibinfo  {journal} {PNAS}\
		}\textbf {\bibinfo {volume} {106}},\ \bibinfo {pages} {10116} (\bibinfo
		{year} {2009})}\BibitemShut {NoStop}%
	\bibitem [{\citenamefont {Kane}\ and\ \citenamefont
		{Fisher}(1994)}]{Kane.1994}%
	\BibitemOpen
	\bibfield  {author} {\bibinfo {author} {\bibfnamefont {C.~L.}\ \bibnamefont
			{Kane}}\ and\ \bibinfo {author} {\bibfnamefont {M.~P.~A.}\ \bibnamefont
			{Fisher}},\ }\href {\doibase 10.1103/PhysRevLett.72.724} {\bibfield
		{journal} {\bibinfo  {journal} {Phys. Rev. Lett.}\ }\textbf {\bibinfo
			{volume} {72}},\ \bibinfo {pages} {724} (\bibinfo {year} {1994})}\BibitemShut
	{NoStop}%
	\bibitem [{\citenamefont {de~Picciotto}\ \emph {et~al.}(1997)\citenamefont
		{de~Picciotto}, \citenamefont {Reznikov}, \citenamefont {Heiblum},
		\citenamefont {Umansky}, \citenamefont {Bunin},\ and\ \citenamefont
		{Mahalu}}]{dePicciotto.1997}%
	\BibitemOpen
	\bibfield  {author} {\bibinfo {author} {\bibfnamefont {R.}~\bibnamefont
			{de~Picciotto}}, \bibinfo {author} {\bibfnamefont {M.}~\bibnamefont
			{Reznikov}}, \bibinfo {author} {\bibfnamefont {M.}~\bibnamefont {Heiblum}},
		\bibinfo {author} {\bibfnamefont {V.}~\bibnamefont {Umansky}}, \bibinfo
		{author} {\bibfnamefont {G.}~\bibnamefont {Bunin}}, \ and\ \bibinfo {author}
		{\bibfnamefont {D.}~\bibnamefont {Mahalu}},\ }\href {\doibase 10.1038/38241}
	{\bibfield  {journal} {\bibinfo  {journal} {Nature}\ }\textbf {\bibinfo
			{volume} {389}},\ \bibinfo {pages} {162} (\bibinfo {year}
		{1997})}\BibitemShut {NoStop}%
	\bibitem [{\citenamefont {Jehl}\ \emph {et~al.}(2000)\citenamefont {Jehl},
		\citenamefont {Sanquer}, \citenamefont {Calemczuk},\ and\ \citenamefont
		{Mailly}}]{Jehl.2000}%
	\BibitemOpen
	\bibfield  {author} {\bibinfo {author} {\bibfnamefont {X.}~\bibnamefont
			{Jehl}}, \bibinfo {author} {\bibfnamefont {M.}~\bibnamefont {Sanquer}},
		\bibinfo {author} {\bibfnamefont {R.}~\bibnamefont {Calemczuk}}, \ and\
		\bibinfo {author} {\bibfnamefont {D.}~\bibnamefont {Mailly}},\ }\href
	{\doibase 10.1038/35011012} {\bibfield  {journal} {\bibinfo  {journal}
			{Nature}\ }\textbf {\bibinfo {volume} {405}},\ \bibinfo {pages} {50}
		(\bibinfo {year} {2000})}\BibitemShut {NoStop}%
	\bibitem [{\citenamefont {Lefloch}\ \emph {et~al.}(2003)\citenamefont
		{Lefloch}, \citenamefont {Hoffmann}, \citenamefont {Sanquer},\ and\
		\citenamefont {Quirion}}]{Lefloch.2003}%
	\BibitemOpen
	\bibfield  {author} {\bibinfo {author} {\bibfnamefont {F.}~\bibnamefont
			{Lefloch}}, \bibinfo {author} {\bibfnamefont {C.}~\bibnamefont {Hoffmann}},
		\bibinfo {author} {\bibfnamefont {M.}~\bibnamefont {Sanquer}}, \ and\
		\bibinfo {author} {\bibfnamefont {D.}~\bibnamefont {Quirion}},\ }\href
	{\doibase 10.1103/PhysRevLett.90.067002} {\bibfield  {journal} {\bibinfo
			{journal} {Phys. Rev. Lett.}\ }\textbf {\bibinfo {volume} {90}},\ \bibinfo
		{pages} {067002} (\bibinfo {year} {2003})}\BibitemShut {NoStop}%
	\bibitem [{\citenamefont {Levitov}\ and\ \citenamefont
		{Lesovik}(1993)}]{Levitov.1993}%
	\BibitemOpen
	\bibfield  {author} {\bibinfo {author} {\bibfnamefont {L.~S.}\ \bibnamefont
			{Levitov}}\ and\ \bibinfo {author} {\bibfnamefont {G.~B.}\ \bibnamefont
			{Lesovik}},\ }\href@noop {} {\bibfield  {journal} {\bibinfo  {journal} {JETP
				Lett.}\ }\textbf {\bibinfo {volume} {58}},\ \bibinfo {pages} {230} (\bibinfo
		{year} {1993})}\BibitemShut {NoStop}%
	\bibitem [{\citenamefont {Levitov}\ \emph {et~al.}(1996)\citenamefont
		{Levitov}, \citenamefont {Lee},\ and\ \citenamefont
		{Lesovik}}]{Levitov.1996}%
	\BibitemOpen
	\bibfield  {author} {\bibinfo {author} {\bibfnamefont {L.~S.}\ \bibnamefont
			{Levitov}}, \bibinfo {author} {\bibfnamefont {H.}~\bibnamefont {Lee}}, \ and\
		\bibinfo {author} {\bibfnamefont {G.~B.}\ \bibnamefont {Lesovik}},\ }\href
	{\doibase 10.1063/1.531672} {\bibfield  {journal} {\bibinfo  {journal} {J.
				Math. Phys.}\ }\textbf {\bibinfo {volume} {37}},\ \bibinfo {pages} {4845}
		(\bibinfo {year} {1996})}\BibitemShut {NoStop}%
	\bibitem [{\citenamefont {Field}\ \emph {et~al.}(1993)\citenamefont {Field},
		\citenamefont {Smith}, \citenamefont {Pepper}, \citenamefont {Ritchie},
		\citenamefont {Frost}, \citenamefont {Jones},\ and\ \citenamefont
		{Hasko}}]{Field.1993}%
	\BibitemOpen
	\bibfield  {author} {\bibinfo {author} {\bibfnamefont {M.}~\bibnamefont
			{Field}}, \bibinfo {author} {\bibfnamefont {C.~G.}\ \bibnamefont {Smith}},
		\bibinfo {author} {\bibfnamefont {M.}~\bibnamefont {Pepper}}, \bibinfo
		{author} {\bibfnamefont {D.~A.}\ \bibnamefont {Ritchie}}, \bibinfo {author}
		{\bibfnamefont {J.~E.~F.}\ \bibnamefont {Frost}}, \bibinfo {author}
		{\bibfnamefont {G.~A.~C.}\ \bibnamefont {Jones}}, \ and\ \bibinfo {author}
		{\bibfnamefont {D.~G.}\ \bibnamefont {Hasko}},\ }\href {\doibase
		10.1103/PhysRevLett.70.1311} {\bibfield  {journal} {\bibinfo  {journal}
			{Phys. Rev. Lett.}\ }\textbf {\bibinfo {volume} {70}},\ \bibinfo {pages}
		{1311} (\bibinfo {year} {1993})}\BibitemShut {NoStop}%
	\bibitem [{\citenamefont {Elzerman}\ \emph {et~al.}(2004)\citenamefont
		{Elzerman}, \citenamefont {Hanson}, \citenamefont {{van Willems Beveren}},
		\citenamefont {Witkamp}, \citenamefont {Vandersypen},\ and\ \citenamefont
		{Kouwenhoven}}]{Elzerman.2004}%
	\BibitemOpen
	\bibfield  {author} {\bibinfo {author} {\bibfnamefont {J.~M.}\ \bibnamefont
			{Elzerman}}, \bibinfo {author} {\bibfnamefont {R.}~\bibnamefont {Hanson}},
		\bibinfo {author} {\bibfnamefont {L.~H.}\ \bibnamefont {{van Willems
					Beveren}}}, \bibinfo {author} {\bibfnamefont {B.}~\bibnamefont {Witkamp}},
		\bibinfo {author} {\bibfnamefont {L.~M.~K.}\ \bibnamefont {Vandersypen}}, \
		and\ \bibinfo {author} {\bibfnamefont {L.~P.}\ \bibnamefont {Kouwenhoven}},\
	}\href {\doibase 10.1038/nature02693} {\bibfield  {journal} {\bibinfo
			{journal} {Nature}\ }\textbf {\bibinfo {volume} {430}},\ \bibinfo {pages}
		{431} (\bibinfo {year} {2004})}\BibitemShut {NoStop}%
	\bibitem [{\citenamefont {Fujisawa}\ \emph {et~al.}(2004)\citenamefont
		{Fujisawa}, \citenamefont {Hayashi}, \citenamefont {Hirayama}, \citenamefont
		{Cheong},\ and\ \citenamefont {Jeong}}]{Fujisawa.2004}%
	\BibitemOpen
	\bibfield  {author} {\bibinfo {author} {\bibfnamefont {T.}~\bibnamefont
			{Fujisawa}}, \bibinfo {author} {\bibfnamefont {T.}~\bibnamefont {Hayashi}},
		\bibinfo {author} {\bibfnamefont {Y.}~\bibnamefont {Hirayama}}, \bibinfo
		{author} {\bibfnamefont {H.~D.}\ \bibnamefont {Cheong}}, \ and\ \bibinfo
		{author} {\bibfnamefont {Y.~H.}\ \bibnamefont {Jeong}},\ }\href
	{https://doi.org/10.1063/1.1691491} {\bibfield  {journal} {\bibinfo
			{journal} {Appl. Phys. Lett.}\ }\textbf {\bibinfo {volume} {84}},\ \bibinfo
		{pages} {2343} (\bibinfo {year} {2004})}\BibitemShut {NoStop}%
	\bibitem [{\citenamefont {Kiyama}\ \emph {et~al.}(2018)\citenamefont {Kiyama},
		\citenamefont {Korsch}, \citenamefont {Nagai}, \citenamefont {Kanai},
		\citenamefont {Matsumoto}, \citenamefont {Hirakawa},\ and\ \citenamefont
		{Oiwa}}]{Kiyama.2018}%
	\BibitemOpen
	\bibfield  {author} {\bibinfo {author} {\bibfnamefont {H.}~\bibnamefont
			{Kiyama}}, \bibinfo {author} {\bibfnamefont {A.}~\bibnamefont {Korsch}},
		\bibinfo {author} {\bibfnamefont {N.}~\bibnamefont {Nagai}}, \bibinfo
		{author} {\bibfnamefont {Y.}~\bibnamefont {Kanai}}, \bibinfo {author}
		{\bibfnamefont {K.}~\bibnamefont {Matsumoto}}, \bibinfo {author}
		{\bibfnamefont {K.}~\bibnamefont {Hirakawa}}, \ and\ \bibinfo {author}
		{\bibfnamefont {A.}~\bibnamefont {Oiwa}},\ }\href
	{https://doi.org/10.1038/s41598-018-31268-x} {\bibfield  {journal} {\bibinfo
			{journal} {Sci. Rep.}\ }\textbf {\bibinfo {volume} {8}},\ \bibinfo {pages}
		{13188} (\bibinfo {year} {2018})}\BibitemShut {NoStop}%
	\bibitem [{\citenamefont {Hofmann}\ \emph {et~al.}(2016)\citenamefont
		{Hofmann}, \citenamefont {Maisi}, \citenamefont {Gold}, \citenamefont
		{Kr{\"a}henmann}, \citenamefont {R{\"o}ssler}, \citenamefont {Basset},
		\citenamefont {M{\"a}rki}, \citenamefont {Reichl}, \citenamefont
		{Wegscheider}, \citenamefont {Ensslin},\ and\ \citenamefont
		{Ihn}}]{Hofmann.2016}%
	\BibitemOpen
	\bibfield  {author} {\bibinfo {author} {\bibfnamefont {A.}~\bibnamefont
			{Hofmann}}, \bibinfo {author} {\bibfnamefont {V.~F.}\ \bibnamefont {Maisi}},
		\bibinfo {author} {\bibfnamefont {C.}~\bibnamefont {Gold}}, \bibinfo {author}
		{\bibfnamefont {T.}~\bibnamefont {Kr{\"a}henmann}}, \bibinfo {author}
		{\bibfnamefont {C.}~\bibnamefont {R{\"o}ssler}}, \bibinfo {author}
		{\bibfnamefont {J.}~\bibnamefont {Basset}}, \bibinfo {author} {\bibfnamefont
			{P.}~\bibnamefont {M{\"a}rki}}, \bibinfo {author} {\bibfnamefont
			{C.}~\bibnamefont {Reichl}}, \bibinfo {author} {\bibfnamefont
			{W.}~\bibnamefont {Wegscheider}}, \bibinfo {author} {\bibfnamefont
			{K.}~\bibnamefont {Ensslin}}, \ and\ \bibinfo {author} {\bibfnamefont
			{T.}~\bibnamefont {Ihn}},\ }\href {\doibase 10.1103/PhysRevLett.117.206803}
	{\bibfield  {journal} {\bibinfo  {journal} {Phys. Rev. Lett.}\ }\textbf
		{\bibinfo {volume} {117}},\ \bibinfo {pages} {206803} (\bibinfo {year}
		{2016})}\BibitemShut {NoStop}%
	\bibitem [{\citenamefont {Maisi}\ \emph {et~al.}(2016)\citenamefont {Maisi},
		\citenamefont {Hofmann}, \citenamefont {R{\"o}{\"o}sli}, \citenamefont
		{Basset}, \citenamefont {Reichl}, \citenamefont {Wegscheider}, \citenamefont
		{Ihn},\ and\ \citenamefont {Ensslin}}]{Maisi.2016}%
	\BibitemOpen
	\bibfield  {author} {\bibinfo {author} {\bibfnamefont {V.~F.}\ \bibnamefont
			{Maisi}}, \bibinfo {author} {\bibfnamefont {A.}~\bibnamefont {Hofmann}},
		\bibinfo {author} {\bibfnamefont {M.}~\bibnamefont {R{\"o}{\"o}sli}},
		\bibinfo {author} {\bibfnamefont {J.}~\bibnamefont {Basset}}, \bibinfo
		{author} {\bibfnamefont {C.}~\bibnamefont {Reichl}}, \bibinfo {author}
		{\bibfnamefont {W.}~\bibnamefont {Wegscheider}}, \bibinfo {author}
		{\bibfnamefont {T.}~\bibnamefont {Ihn}}, \ and\ \bibinfo {author}
		{\bibfnamefont {K.}~\bibnamefont {Ensslin}},\ }\href {\doibase
		10.1103/PhysRevLett.116.136803} {\bibfield  {journal} {\bibinfo  {journal}
			{Phys. Rev. Lett.}\ }\textbf {\bibinfo {volume} {116}},\ \bibinfo {pages}
		{136803} (\bibinfo {year} {2016})}\BibitemShut {NoStop}%
	\bibitem [{\citenamefont {Hofmann}\ \emph {et~al.}(2017)\citenamefont
		{Hofmann}, \citenamefont {Maisi}, \citenamefont {Kr{\"a}henmann},
		\citenamefont {Reichl}, \citenamefont {Wegscheider}, \citenamefont
		{Ensslin},\ and\ \citenamefont {Ihn}}]{Hofmann.2017}%
	\BibitemOpen
	\bibfield  {author} {\bibinfo {author} {\bibfnamefont {A.}~\bibnamefont
			{Hofmann}}, \bibinfo {author} {\bibfnamefont {V.~F.}\ \bibnamefont {Maisi}},
		\bibinfo {author} {\bibfnamefont {T.}~\bibnamefont {Kr{\"a}henmann}},
		\bibinfo {author} {\bibfnamefont {C.}~\bibnamefont {Reichl}}, \bibinfo
		{author} {\bibfnamefont {W.}~\bibnamefont {Wegscheider}}, \bibinfo {author}
		{\bibfnamefont {K.}~\bibnamefont {Ensslin}}, \ and\ \bibinfo {author}
		{\bibfnamefont {T.}~\bibnamefont {Ihn}},\ }\href {\doibase
		10.1103/PhysRevLett.119.176807} {\bibfield  {journal} {\bibinfo  {journal}
			{Phys. Rev. Lett.}\ }\textbf {\bibinfo {volume} {119}},\ \bibinfo {pages}
		{176807} (\bibinfo {year} {2017})}\BibitemShut {NoStop}%
	\bibitem [{\citenamefont {Vamivakas}\ \emph {et~al.}(2010)\citenamefont
		{Vamivakas}, \citenamefont {Lu}, \citenamefont {Matthiesen}, \citenamefont
		{Zhao}, \citenamefont {F{\"a}lt}, \citenamefont {Badolato},\ and\
		\citenamefont {Atat{\"u}re}}]{Vamivakas.2010}%
	\BibitemOpen
	\bibfield  {author} {\bibinfo {author} {\bibfnamefont {A.~N.}\ \bibnamefont
			{Vamivakas}}, \bibinfo {author} {\bibfnamefont {C.-Y.}\ \bibnamefont {Lu}},
		\bibinfo {author} {\bibfnamefont {C.}~\bibnamefont {Matthiesen}}, \bibinfo
		{author} {\bibfnamefont {Y.}~\bibnamefont {Zhao}}, \bibinfo {author}
		{\bibfnamefont {S.}~\bibnamefont {F{\"a}lt}}, \bibinfo {author}
		{\bibfnamefont {A.}~\bibnamefont {Badolato}}, \ and\ \bibinfo {author}
		{\bibfnamefont {M.}~\bibnamefont {Atat{\"u}re}},\ }\href {\doibase
		10.1038/nature09359} {\bibfield  {journal} {\bibinfo  {journal} {Nature}\
		}\textbf {\bibinfo {volume} {467}},\ \bibinfo {pages} {297} (\bibinfo {year}
		{2010})}\BibitemShut {NoStop}%
	\bibitem [{\citenamefont {Matthiesen}\ \emph {et~al.}(2013)\citenamefont
		{Matthiesen}, \citenamefont {Geller}, \citenamefont {Schulte}, \citenamefont
		{{Le Gall}}, \citenamefont {Hansom}, \citenamefont {Li}, \citenamefont
		{Hugues}, \citenamefont {Clarke},\ and\ \citenamefont
		{Atat{\"u}re}}]{Matthiesen.2013}%
	\BibitemOpen
	\bibfield  {author} {\bibinfo {author} {\bibfnamefont {C.}~\bibnamefont
			{Matthiesen}}, \bibinfo {author} {\bibfnamefont {M.}~\bibnamefont {Geller}},
		\bibinfo {author} {\bibfnamefont {C.~H.~H.}\ \bibnamefont {Schulte}},
		\bibinfo {author} {\bibfnamefont {C.}~\bibnamefont {{Le Gall}}}, \bibinfo
		{author} {\bibfnamefont {J.}~\bibnamefont {Hansom}}, \bibinfo {author}
		{\bibfnamefont {Z.}~\bibnamefont {Li}}, \bibinfo {author} {\bibfnamefont
			{M.}~\bibnamefont {Hugues}}, \bibinfo {author} {\bibfnamefont
			{E.}~\bibnamefont {Clarke}}, \ and\ \bibinfo {author} {\bibfnamefont
			{M.}~\bibnamefont {Atat{\"u}re}},\ }\href {\doibase 10.1038/ncomms2601}
	{\bibfield  {journal} {\bibinfo  {journal} {Nature Commun.}\ }\textbf
		{\bibinfo {volume} {4}},\ \bibinfo {pages} {1600} (\bibinfo {year}
		{2013})}\BibitemShut {NoStop}%
	\bibitem [{\citenamefont {Kurzmann}\ \emph
		{et~al.}(2016{\natexlab{a}})\citenamefont {Kurzmann}, \citenamefont {Merkel},
		\citenamefont {Labud}, \citenamefont {Ludwig}, \citenamefont {Wieck},
		\citenamefont {Lorke},\ and\ \citenamefont {Geller}}]{Kurzmann.2016b}%
	\BibitemOpen
	\bibfield  {author} {\bibinfo {author} {\bibfnamefont {A.}~\bibnamefont
			{Kurzmann}}, \bibinfo {author} {\bibfnamefont {B.}~\bibnamefont {Merkel}},
		\bibinfo {author} {\bibfnamefont {P.~A.}\ \bibnamefont {Labud}}, \bibinfo
		{author} {\bibfnamefont {A.}~\bibnamefont {Ludwig}}, \bibinfo {author}
		{\bibfnamefont {A.~D.}\ \bibnamefont {Wieck}}, \bibinfo {author}
		{\bibfnamefont {A.}~\bibnamefont {Lorke}}, \ and\ \bibinfo {author}
		{\bibfnamefont {M.}~\bibnamefont {Geller}},\ }\href {\doibase
		10.1103/PhysRevLett.117.017401} {\bibfield  {journal} {\bibinfo  {journal}
			{Phys. Rev. Lett.}\ }\textbf {\bibinfo {volume} {117}},\ \bibinfo {pages}
		{017401} (\bibinfo {year} {2016}{\natexlab{a}})}\BibitemShut {NoStop}%
	\bibitem [{\citenamefont {Kurzmann}\ \emph
		{et~al.}(2016{\natexlab{b}})\citenamefont {Kurzmann}, \citenamefont {Ludwig},
		\citenamefont {Wieck}, \citenamefont {Lorke},\ and\ \citenamefont
		{Geller}}]{Kurzmann.2016c}%
	\BibitemOpen
	\bibfield  {author} {\bibinfo {author} {\bibfnamefont {A.}~\bibnamefont
			{Kurzmann}}, \bibinfo {author} {\bibfnamefont {A.}~\bibnamefont {Ludwig}},
		\bibinfo {author} {\bibfnamefont {A.~D.}\ \bibnamefont {Wieck}}, \bibinfo
		{author} {\bibfnamefont {A.}~\bibnamefont {Lorke}}, \ and\ \bibinfo {author}
		{\bibfnamefont {M.}~\bibnamefont {Geller}},\ }\href {\doibase
		10.1063/1.4954944} {\bibfield  {journal} {\bibinfo  {journal} {Appl. Phys.
				Lett.}\ }\textbf {\bibinfo {volume} {108}},\ \bibinfo {pages} {263108}
		(\bibinfo {year} {2016}{\natexlab{b}})}\BibitemShut {NoStop}%
	\bibitem [{PRL(sign)}]{PRLSuppl1.2019}%
	\BibitemOpen
	\href@noop {} {} \bibinfo {year} {See Supplemental Material at [Url], which
		includes Refs. [36-40], for sample and device design}\BibitemShut {NoStop}%
	\bibitem [{PRL(ment)}]{PRLSuppl2.2019}%
	\BibitemOpen
	\href@noop {} {} \bibinfo {year} {See Supplemental Material at [Url], which
		includes Refs. [36-40], for more details on the optical
		measurement}\BibitemShut {NoStop}%
	\bibitem [{\citenamefont {Kuhlmann}\ \emph {et~al.}(2013)\citenamefont
		{Kuhlmann}, \citenamefont {Houel}, \citenamefont {Ludwig}, \citenamefont
		{Greuter}, \citenamefont {Reuter}, \citenamefont {Wieck}, \citenamefont
		{Poggio},\ and\ \citenamefont {Warburton}}]{Kuhlmann.2013}%
	\BibitemOpen
	\bibfield  {author} {\bibinfo {author} {\bibfnamefont {A.~V.}\ \bibnamefont
			{Kuhlmann}}, \bibinfo {author} {\bibfnamefont {J.}~\bibnamefont {Houel}},
		\bibinfo {author} {\bibfnamefont {A.}~\bibnamefont {Ludwig}}, \bibinfo
		{author} {\bibfnamefont {L.}~\bibnamefont {Greuter}}, \bibinfo {author}
		{\bibfnamefont {D.}~\bibnamefont {Reuter}}, \bibinfo {author} {\bibfnamefont
			{A.~D.}\ \bibnamefont {Wieck}}, \bibinfo {author} {\bibfnamefont
			{M.}~\bibnamefont {Poggio}}, \ and\ \bibinfo {author} {\bibfnamefont {R.~J.}\
			\bibnamefont {Warburton}},\ }\href {\doibase 10.1038/nphys2688} {\bibfield
		{journal} {\bibinfo  {journal} {Nature Phys.}\ }\textbf {\bibinfo {volume}
			{9}},\ \bibinfo {pages} {570} (\bibinfo {year} {2013})}\BibitemShut {NoStop}%
	\bibitem [{\citenamefont {Latta}\ \emph {et~al.}(2009)\citenamefont {Latta},
		\citenamefont {H{\"o}gele}, \citenamefont {Zhao}, \citenamefont {Vamivakas},
		\citenamefont {Maletinsky}, \citenamefont {Kroner}, \citenamefont {Dreiser},
		\citenamefont {Carusotto}, \citenamefont {Badolato}, \citenamefont {Schuh},
		\citenamefont {Wegscheider}, \citenamefont {Atat{\"u}re},\ and\ \citenamefont
		{Imamo{\u{g}}lu}}]{Latta.2009}%
	\BibitemOpen
	\bibfield  {author} {\bibinfo {author} {\bibfnamefont {C.}~\bibnamefont
			{Latta}}, \bibinfo {author} {\bibfnamefont {A.}~\bibnamefont {H{\"o}gele}},
		\bibinfo {author} {\bibfnamefont {Y.}~\bibnamefont {Zhao}}, \bibinfo {author}
		{\bibfnamefont {A.~N.}\ \bibnamefont {Vamivakas}}, \bibinfo {author}
		{\bibfnamefont {P.}~\bibnamefont {Maletinsky}}, \bibinfo {author}
		{\bibfnamefont {M.}~\bibnamefont {Kroner}}, \bibinfo {author} {\bibfnamefont
			{J.}~\bibnamefont {Dreiser}}, \bibinfo {author} {\bibfnamefont
			{I.}~\bibnamefont {Carusotto}}, \bibinfo {author} {\bibfnamefont
			{A.}~\bibnamefont {Badolato}}, \bibinfo {author} {\bibfnamefont
			{D.}~\bibnamefont {Schuh}}, \bibinfo {author} {\bibfnamefont
			{W.}~\bibnamefont {Wegscheider}}, \bibinfo {author} {\bibfnamefont
			{M.}~\bibnamefont {Atat{\"u}re}}, \ and\ \bibinfo {author} {\bibfnamefont
			{A.}~\bibnamefont {Imamo{\u{g}}lu}},\ }\href {\doibase 10.1038/nphys1363}
	{\bibfield  {journal} {\bibinfo  {journal} {Nature Phys.}\ }\textbf {\bibinfo
			{volume} {5}},\ \bibinfo {pages} {758} (\bibinfo {year} {2009})}\BibitemShut
	{NoStop}%
	\bibitem [{\citenamefont {H{\"o}gele}\ \emph {et~al.}(2012)\citenamefont
		{H{\"o}gele}, \citenamefont {Kroner}, \citenamefont {Latta}, \citenamefont
		{Claassen}, \citenamefont {Carusotto}, \citenamefont {Bulutay},\ and\
		\citenamefont {Imamo{\u{g}}lu}}]{Hogele.2012}%
	\BibitemOpen
	\bibfield  {author} {\bibinfo {author} {\bibfnamefont {A.}~\bibnamefont
			{H{\"o}gele}}, \bibinfo {author} {\bibfnamefont {M.}~\bibnamefont {Kroner}},
		\bibinfo {author} {\bibfnamefont {C.}~\bibnamefont {Latta}}, \bibinfo
		{author} {\bibfnamefont {M.}~\bibnamefont {Claassen}}, \bibinfo {author}
		{\bibfnamefont {I.}~\bibnamefont {Carusotto}}, \bibinfo {author}
		{\bibfnamefont {C.}~\bibnamefont {Bulutay}}, \ and\ \bibinfo {author}
		{\bibfnamefont {A.}~\bibnamefont {Imamo{\u{g}}lu}},\ }\href {\doibase
		10.1103/PhysRevLett.108.197403} {\bibfield  {journal} {\bibinfo  {journal}
			{Phys. Rev. Lett.}\ }\textbf {\bibinfo {volume} {108}},\ \bibinfo {pages}
		{197403} (\bibinfo {year} {2012})}\BibitemShut {NoStop}%
	\bibitem [{PRL(stem)}]{PRLSuppl3.2019}%
	\BibitemOpen
	\href@noop {} {} \bibinfo {year} {See Supplemental Material at [Url], which
		includes Refs. [36-40], for more details on the model system}\BibitemShut
	{NoStop}%
	\bibitem [{\citenamefont {Hanson}\ \emph {et~al.}(2005)\citenamefont {Hanson},
		\citenamefont {{van Beveren}}, \citenamefont {Vink}, \citenamefont
		{Elzerman}, \citenamefont {Naber}, \citenamefont {Koppens}, \citenamefont
		{Kouwenhoven},\ and\ \citenamefont {Vandersypen}}]{Hanson.2005}%
	\BibitemOpen
	\bibfield  {author} {\bibinfo {author} {\bibfnamefont {R.}~\bibnamefont
			{Hanson}}, \bibinfo {author} {\bibfnamefont {L.~H.~W.}\ \bibnamefont {{van
					Beveren}}}, \bibinfo {author} {\bibfnamefont {I.~T.}\ \bibnamefont {Vink}},
		\bibinfo {author} {\bibfnamefont {J.~M.}\ \bibnamefont {Elzerman}}, \bibinfo
		{author} {\bibfnamefont {W.~J.~M.}\ \bibnamefont {Naber}}, \bibinfo {author}
		{\bibfnamefont {F.~H.~L.}\ \bibnamefont {Koppens}}, \bibinfo {author}
		{\bibfnamefont {L.~P.}\ \bibnamefont {Kouwenhoven}}, \ and\ \bibinfo {author}
		{\bibfnamefont {L.~M.~K.}\ \bibnamefont {Vandersypen}},\ }\href {\doibase
		10.1103/PhysRevLett.94.196802} {\bibfield  {journal} {\bibinfo  {journal}
			{Phys. Rev. Lett.}\ }\textbf {\bibinfo {volume} {94}},\ \bibinfo {pages}
		{196802} (\bibinfo {year} {2005})}\BibitemShut {NoStop}%
	\bibitem [{\citenamefont {Kroutvar}\ \emph {et~al.}(2004)\citenamefont
		{Kroutvar}, \citenamefont {Ducommun}, \citenamefont {Heiss}, \citenamefont
		{Bichler}, \citenamefont {Schuh}, \citenamefont {Abstreiter},\ and\
		\citenamefont {Finley}}]{Kroutvar.2004}%
	\BibitemOpen
	\bibfield  {author} {\bibinfo {author} {\bibfnamefont {M.}~\bibnamefont
			{Kroutvar}}, \bibinfo {author} {\bibfnamefont {Y.}~\bibnamefont {Ducommun}},
		\bibinfo {author} {\bibfnamefont {D.}~\bibnamefont {Heiss}}, \bibinfo
		{author} {\bibfnamefont {M.}~\bibnamefont {Bichler}}, \bibinfo {author}
		{\bibfnamefont {D.}~\bibnamefont {Schuh}}, \bibinfo {author} {\bibfnamefont
			{G.}~\bibnamefont {Abstreiter}}, \ and\ \bibinfo {author} {\bibfnamefont
			{J.~J.}\ \bibnamefont {Finley}},\ }\href {\doibase 10.1038/nature03008}
	{\bibfield  {journal} {\bibinfo  {journal} {Nature}\ }\textbf {\bibinfo
			{volume} {432}},\ \bibinfo {pages} {81} (\bibinfo {year} {2004})}\BibitemShut
	{NoStop}%
	\bibitem [{\citenamefont {Kambly}\ \emph {et~al.}(2011)\citenamefont {Kambly},
		\citenamefont {Flindt},\ and\ \citenamefont {B{\"u}ttiker}}]{Kambly.2011}%
	\BibitemOpen
	\bibfield  {author} {\bibinfo {author} {\bibfnamefont {D.}~\bibnamefont
			{Kambly}}, \bibinfo {author} {\bibfnamefont {C.}~\bibnamefont {Flindt}}, \
		and\ \bibinfo {author} {\bibfnamefont {M.}~\bibnamefont {B{\"u}ttiker}},\
	}\href {https://doi.org/10.1103/PhysRevB.83.075432} {\bibfield  {journal}
		{\bibinfo  {journal} {Phys. Rev. B}\ }\textbf {\bibinfo {volume} {83}},\
		\bibinfo {pages} {230} (\bibinfo {year} {2011})}\BibitemShut {NoStop}%
	\bibitem [{\citenamefont {Stegmann}\ \emph {et~al.}(2015)\citenamefont
		{Stegmann}, \citenamefont {Sothmann}, \citenamefont {Hucht},\ and\
		\citenamefont {K{\"o}nig}}]{Stegmann.2015}%
	\BibitemOpen
	\bibfield  {author} {\bibinfo {author} {\bibfnamefont {P.}~\bibnamefont
			{Stegmann}}, \bibinfo {author} {\bibfnamefont {B.}~\bibnamefont {Sothmann}},
		\bibinfo {author} {\bibfnamefont {A.}~\bibnamefont {Hucht}}, \ and\ \bibinfo
		{author} {\bibfnamefont {J.}~\bibnamefont {K{\"o}nig}},\ }\href
	{https://doi.org/10.1103/PhysRevB.92.155413} {\bibfield  {journal} {\bibinfo
			{journal} {Phys. Rev. B}\ }\textbf {\bibinfo {volume} {92}},\ \bibinfo
		{pages} {592} (\bibinfo {year} {2015})}\BibitemShut {NoStop}%
	\bibitem [{\citenamefont {Stegmann}\ and\ \citenamefont
		{K{\"o}nig}(2016)}]{Stegmann.2016}%
	\BibitemOpen
	\bibfield  {author} {\bibinfo {author} {\bibfnamefont {P.}~\bibnamefont
			{Stegmann}}\ and\ \bibinfo {author} {\bibfnamefont {J.}~\bibnamefont
			{K{\"o}nig}},\ }\href {http://link.aps.org/pdf/10.1103/PhysRevB.94.125433}
	{\bibfield  {journal} {\bibinfo  {journal} {Phys. Rev. B}\ }\textbf {\bibinfo
			{volume} {94}},\ \bibinfo {pages} {125433} (\bibinfo {year}
		{2016})}\BibitemShut {NoStop}%
	\bibitem [{\citenamefont {Fricke}\ \emph {et~al.}(2010)\citenamefont {Fricke},
		\citenamefont {Hohls}, \citenamefont {Sethubalasubramanian}, \citenamefont
		{Fricke},\ and\ \citenamefont {Haug}}]{Fricke.2010}%
	\BibitemOpen
	\bibfield  {author} {\bibinfo {author} {\bibfnamefont {C.}~\bibnamefont
			{Fricke}}, \bibinfo {author} {\bibfnamefont {F.}~\bibnamefont {Hohls}},
		\bibinfo {author} {\bibfnamefont {N.}~\bibnamefont {Sethubalasubramanian}},
		\bibinfo {author} {\bibfnamefont {L.}~\bibnamefont {Fricke}}, \ and\ \bibinfo
		{author} {\bibfnamefont {R.~J.}\ \bibnamefont {Haug}},\ }\href
	{https://doi.org/10.1063/1.3430000} {\bibfield  {journal} {\bibinfo
			{journal} {Appl. Phys. Lett.}\ }\textbf {\bibinfo {volume} {96}},\ \bibinfo
		{pages} {202103} (\bibinfo {year} {2010})}\BibitemShut {NoStop}%
	\bibitem [{\citenamefont {Flagg}\ \emph {et~al.}(2009)\citenamefont {Flagg},
		\citenamefont {Muller}, \citenamefont {Robertson}, \citenamefont {Founta},
		\citenamefont {Deppe}, \citenamefont {Xiao}, \citenamefont {Ma},
		\citenamefont {Salamo},\ and\ \citenamefont {Shih}}]{Flagg.2009}%
	\BibitemOpen
	\bibfield  {author} {\bibinfo {author} {\bibfnamefont {E.~B.}\ \bibnamefont
			{Flagg}}, \bibinfo {author} {\bibfnamefont {A.}~\bibnamefont {Muller}},
		\bibinfo {author} {\bibfnamefont {J.~W.}\ \bibnamefont {Robertson}}, \bibinfo
		{author} {\bibfnamefont {S.}~\bibnamefont {Founta}}, \bibinfo {author}
		{\bibfnamefont {D.~G.}\ \bibnamefont {Deppe}}, \bibinfo {author}
		{\bibfnamefont {M.}~\bibnamefont {Xiao}}, \bibinfo {author} {\bibfnamefont
			{W.}~\bibnamefont {Ma}}, \bibinfo {author} {\bibfnamefont {G.~J.}\
			\bibnamefont {Salamo}}, \ and\ \bibinfo {author} {\bibfnamefont {C.~K.}\
			\bibnamefont {Shih}},\ }\href {\doibase 10.1038/nphys1184} {\bibfield
		{journal} {\bibinfo  {journal} {Nature Phys.}\ }\textbf {\bibinfo {volume}
			{5}},\ \bibinfo {pages} {203} (\bibinfo {year} {2009})}\BibitemShut {NoStop}%
	\bibitem [{\citenamefont {Muller}\ \emph {et~al.}(2007)\citenamefont {Muller},
		\citenamefont {Flagg}, \citenamefont {Bianucci}, \citenamefont {Wang},
		\citenamefont {Deppe}, \citenamefont {Ma}, \citenamefont {Zhang},
		\citenamefont {Salamo}, \citenamefont {Xiao},\ and\ \citenamefont
		{Shih}}]{Muller.2007}%
	\BibitemOpen
	\bibfield  {author} {\bibinfo {author} {\bibfnamefont {A.}~\bibnamefont
			{Muller}}, \bibinfo {author} {\bibfnamefont {E.~B.}\ \bibnamefont {Flagg}},
		\bibinfo {author} {\bibfnamefont {P.}~\bibnamefont {Bianucci}}, \bibinfo
		{author} {\bibfnamefont {X.~Y.}\ \bibnamefont {Wang}}, \bibinfo {author}
		{\bibfnamefont {D.~G.}\ \bibnamefont {Deppe}}, \bibinfo {author}
		{\bibfnamefont {W.}~\bibnamefont {Ma}}, \bibinfo {author} {\bibfnamefont
			{J.}~\bibnamefont {Zhang}}, \bibinfo {author} {\bibfnamefont {G.~J.}\
			\bibnamefont {Salamo}}, \bibinfo {author} {\bibfnamefont {M.}~\bibnamefont
			{Xiao}}, \ and\ \bibinfo {author} {\bibfnamefont {C.~K.}\ \bibnamefont
			{Shih}},\ }\href {\doibase 10.1103/PhysRevLett.99.187402} {\bibfield
		{journal} {\bibinfo  {journal} {Phys. Rev. Lett.}\ }\textbf {\bibinfo
			{volume} {99}},\ \bibinfo {pages} {187402} (\bibinfo {year}
		{2007})}\BibitemShut {NoStop}%
	\bibitem [{\citenamefont {Loudon}(2010)}]{Loudon.2010}%
	\BibitemOpen
	\bibfield  {author} {\bibinfo {author} {\bibfnamefont {R.}~\bibnamefont
			{Loudon}},\ }\href@noop {} {\emph {\bibinfo {title} {The quantum theory of
				light}}},\ \bibinfo {edition} {3rd}\ ed.,\ Oxford science publications\
	(\bibinfo  {publisher} {{Oxford Univ. Press}},\ \bibinfo {address} {Oxford},\
	\bibinfo {year} {2010})\BibitemShut {NoStop}%
	\bibitem [{\citenamefont {Bayer}\ \emph {et~al.}(2002)\citenamefont {Bayer},
		\citenamefont {Ortner}, \citenamefont {Stern}, \citenamefont {Kuther},
		\citenamefont {Gorbunov}, \citenamefont {Forchel}, \citenamefont {Hawrylak},
		\citenamefont {Fafard}, \citenamefont {Hinzer}, \citenamefont {Reinecke},
		\citenamefont {Walck}, \citenamefont {Reithmaier}, \citenamefont {Klopf},\
		and\ \citenamefont {Sch{\"a}fer}}]{Bayer.2002}%
	\BibitemOpen
	\bibfield  {author} {\bibinfo {author} {\bibfnamefont {M.}~\bibnamefont
			{Bayer}}, \bibinfo {author} {\bibfnamefont {G.}~\bibnamefont {Ortner}},
		\bibinfo {author} {\bibfnamefont {O.}~\bibnamefont {Stern}}, \bibinfo
		{author} {\bibfnamefont {A.}~\bibnamefont {Kuther}}, \bibinfo {author}
		{\bibfnamefont {A.~A.}\ \bibnamefont {Gorbunov}}, \bibinfo {author}
		{\bibfnamefont {A.}~\bibnamefont {Forchel}}, \bibinfo {author} {\bibfnamefont
			{P.}~\bibnamefont {Hawrylak}}, \bibinfo {author} {\bibfnamefont
			{S.}~\bibnamefont {Fafard}}, \bibinfo {author} {\bibfnamefont
			{K.}~\bibnamefont {Hinzer}}, \bibinfo {author} {\bibfnamefont {T.~L.}\
			\bibnamefont {Reinecke}}, \bibinfo {author} {\bibfnamefont {S.~N.}\
			\bibnamefont {Walck}}, \bibinfo {author} {\bibfnamefont {J.~P.}\ \bibnamefont
			{Reithmaier}}, \bibinfo {author} {\bibfnamefont {F.}~\bibnamefont {Klopf}}, \
		and\ \bibinfo {author} {\bibfnamefont {F.}~\bibnamefont {Sch{\"a}fer}},\
	}\href {https://journals.aps.org/prb/abstract/10.1103/PhysRevB.65.195315}
	{\bibfield  {journal} {\bibinfo  {journal} {Phys. Rev. B}\ }\textbf {\bibinfo
			{volume} {65}} (\bibinfo {year} {2002})}\BibitemShut {NoStop}%
	\bibitem [{\citenamefont {Kleinherbers}\ \emph {et~al.}(2018)\citenamefont
		{Kleinherbers}, \citenamefont {Stegmann},\ and\ \citenamefont
		{K{\"o}nig}}]{Kleinherbers.2018}%
	\BibitemOpen
	\bibfield  {author} {\bibinfo {author} {\bibfnamefont {E.}~\bibnamefont
			{Kleinherbers}}, \bibinfo {author} {\bibfnamefont {P.}~\bibnamefont
			{Stegmann}}, \ and\ \bibinfo {author} {\bibfnamefont {J.}~\bibnamefont
			{K{\"o}nig}},\ }\href {\doibase 10.1088/1367-2630/aad14a} {\bibfield
		{journal} {\bibinfo  {journal} {New J. Phys.}\ }\textbf {\bibinfo {volume}
			{20}},\ \bibinfo {pages} {073023} (\bibinfo {year} {2018})}\BibitemShut
	{NoStop}%
\end{thebibliography}

%

\end{document}